# NEAR-INFRARED CIRCULAR POLARIMETRY AND CORRELATION DIAGRAMS IN THE ORION BN/KL REGION: CONTRIBUTION OF DICHROIC EXTINCTION

Short title: NIR CIRCULAR POLARIMETRY OF THE BN/KL REGION

T. FUKUE,[1] M. TAMURA,[2,3] R. KANDORI,[2] N. KUSAKABE,[2] J. H. HOUGH,[4] P. W. LUCAS,[4] J. BAILEY,[5] D. C. B. WHITTET,[6] Y. NAKAJIMA,[2] J. HASHIMOTO,[3] AND T. NAGATA[1]

[1] Department of Astronomy, Kyoto University, Sakyo-ku, Kyoto 606-8502, Japan; tsubasa@kusastro.kyoto-u.ac.jp.

[2] National Astronomical Observatory, 2-21-1 Osawa, Mitaka, Tokyo 181-8588, Japan.

[3] Graduate University of Advanced Science, 2-21-1 Osawa, Mitaka, Tokyo 181-8588, Japan.

[4] Centre for Astrophysics Research, Science and Technology Research Institute, University of Hertfordshire, Hatfield, Herts. AL10 9AB, UK.

[5] School of Physics, University of New South Wales, NSW 2052, Australia.

[6] Department of Physics and Astronomy, Rensselaer Polytechnic Institute, Troy, NY 12180-3590, USA.

ABSTRACT

We present a deep circular polarization image of the Orion BN/KL nebula in the $K_s$ band and correlations of circular polarization, linear polarization, and $H$-$K_s$ color representing extinction. The image of circular polarization clearly reveals the quadrupolar structure around the massive star IRc2, rather than BN. $H$-$K_s$ color is well correlated with circular polarization. A simple relation between dichroic extinction, color excess, circular and linear polarization is derived. The observed correlation between the Stokes parameters and the color excess agrees with the derived relation, and suggests a major contribution of dichroic extinction to the production of circular polarization in this region, indicating the wide existence of aligned grains.

*Subject headings:* circumstellar matter—infrared: stars—ISM: individual (Orion Kleinmann-Low)—dust, extinction—polarization—stars: formation

# 1. INTRODUCTION

Near-infrared linear and circular polarimetry of star-forming regions can provide detailed information on the optical and physical properties of the dust grains. Circular polarization (CP) is influenced by the grain alignment and the system inclination (Whitney & Wolff 2002; Lucas et al. 2004, 2005). When the grains are aligned with the magnetic field (Lazarian 2007) and dichroic extinction is effective, circular polarimetry can reveal the field structure, e.g., the helical magnetic field detected by Chrysostomou et al. (2007). There are three likely mechanisms for CP in star forming regions: multiple scattering in an optically thick region of dust grains, dichroic scattering by (partially) aligned non-spherical dust grains, and dichroic extinction of linearly polarized radiation in (partially) aligned non-spherical dust grains (Ménard et al. 1988; Chrysostomou et al. 2000; Gledhill & McCall 2000; Buschermöhle et al. 2005). Although the effective mechanism of CP may depend on the environment, previous studies show that higher mass young stellar objects have larger CP (Gledhill et al. 1996; Bailey et al. 1998; Clark et al. 2000; Ménard et al. 2000; Clayton et al. 2005; Chrysostomou et al. 2007). Further studies of star-forming regions by near-infrared imaging circular polarimetry are necessary to assess the generation mechanism for CP.

The massive star-forming region of the Becklin-Neugebauer/Kleinman-Low (BN/KL) nebula has relatively high and widespread CP with a quadrupolar pattern (Bailey et al. 1998; Chrysostomou et al. 2000). It is a very good nearby site to investigate the CP mechanisms. Near the center of the high CP region are the massive protostars IRc2 and BN, with masses of 25 and >7 $M_{sun}$, respectively (Genzel & Stutzki 1989). Chrysostomou et al.

(2000) show a distinct correlation between high degrees of CP and linear polarization (LP). They argue the importance of dichroic scattering off aligned grains. On the other hand, Buschermöhle et al. (2005) suggests the importance of the dichroic extinction in aligned non-spherical grains for CP. They show the comparison of *J-K* color and CP for sample strips in the region, indicating the correlation between extinction and CP.

In this paper, we present a deep near-infrared CP image of the BN/KL region. Moreover, the observed correlations between the Stokes parameters and the color excess (a proxy for extinction) are also presented. To assess the generation mechanism for CP, a relation between CP, color excess and the position angle of LP is derived and compared with the observations.

## 2. OBSERVATIONS AND DATA REDUCTION

The 2.14 μm ($K_s$ band) imaging circular polarimetry data of M42 were obtained with the SIRIUS camera (Nagayama et al. 2003) and its polarimeter on the 1.4-m IRSF telescope at the South African Astronomical Observatory, on nights during 2006 December. The SIRIUS camera also produces 1.25 μm (*J* band) and 1.63 μm (*H* band) images simultaneously. For the circular polarimetry measurements, a polarimeter unit consisting of achromatic half- and quarter-wave plates and a fixed high-extinction-ratio wire-grid analyzer was attached to SIRIUS. For the linear polarimetry measurements and details of the instruments used, see Tamura et al. (2006), Kandori et al. (2006), and references therein.

The total integration time per wave plate position was 900 seconds, and the resultant seeing size was about 1.5″. One pixel corresponds to 0.45″. Exposures were taken at two position angles (P.A.'s) of the quarter-wave plate, in the sequence of P.A. = 0° and 90°. The uncertainty of polarization measurements due to the stability of the sky was estimated to be about 0.3% (Kandori et al. 2006). To remove possible linear-to-circular polarization conversion by the instrument, the half-wave plate in front of the quarter-wave plate was continuously rotated during circular polarimetry measurements (Chrysostomou et al. 1997). The instrumental CP was confirmed to be less than ~0.3% when ~100 % LP was incident.

After image calibrations in the standard manner using IRAF (dark subtraction, flat-fielding with twilight-flats, and sky subtraction), the Stokes parameter images were obtained using $V = I$(P.A. = 0°) - $I$(P.A. = 90°), and $I = I$(P.A. = 0°) + $I$(P.A. = 90°). The CP degree images are then derived using $P = V / I$. Furthermore, the CP image was Gaussian-smoothed with a FWHM of three pixels to match the seeing size of the observations.

The smoothed data are used to construct pixel-by-pixel correlation diagrams (Figs. 2 and 3), where each axis is divided into 100 bins, and the number (indicated by $N$) of pixels whose values go into the two-dimensional bins is shown by a logarithmic color bar in the diagrams. Surface brightness can be computed by MAG (mag arcsec$^{-2}$) = ZMAG - 2.5 $\log_{10}$(FLUX) + 2.5 $\log_{10}$(0.45$^2$), where ZMAG of 20.70 and 19.97 for the $H$ and $K_s$ band is the zero point for magnitudes calibrated by using 2MASS-PSC (Cutri et al. 2003). FLUX denotes the flux measured at each pixel with a scale of 0.45″ pixel$^{-1}$. The 10σ limiting magnitude for surface brightness is $H$ = 19.3 and $K_s$ = 18.0 mag arcsec$^{-2}$ after smoothing images with a Gaussian kernel to match the seeing size (~1.5″).

The estimated uncertainties in polarization distributions are less than about 0.3% for both LP and CP images except for the regions around the north-side corners of the CP image where the uncertainties reach about 0.6 %. Ghost images caused by the polarimeter optics appear as faint circular patterns centered on the Trapezium and its south in the CP image. In this study, we focus on the observed central region with a field-of-view of 2.7′ square around the BN/KL nebula having significant CP. In this region, the correlations studied here are not affected by the ghosts.

## 3. RESULTS AND DISCUSSION

### 3.1. *Quadrupolar Structure of the CP Image around the Massive Star*

Figure 1*a* shows the image of the degree of CP in the $K_s$ band of the central region of the Orion star-forming region. The degree of CP ranges from +17% to −5%. Bright point-like sources are not completely cancelled and are visible in this image even if they are unpolarized, due to slight misalignment among different images. Figure 1*b* shows the *H-$K_s$* color image overlaid with contours of the corresponding CP degree in the $K_s$ band. The image indicates that CP is well-correlated with extinction; highly obscured regions with large *H-$K_s$* correspond to regions having large absolute CP.

In Figure 1, the handedness of CP along the east-west axis is clearly opposite to that along the north-south axis, revealing an asymmetric quadrupolar pattern of CP in this region (Chrysostomou et al. 2000). The LP pattern observed at 2.2 μm is centrosymmetric around two illuminating sources: IRc2 and BN (Minchin et al. 1991; Tamura et al. 2006), so it is more complicated than the pattern associated with a single source. In this paper, the deep CP image in Figure 1 clearly indicates that the quadrupolar pattern is centered on IRc2 (indicated by arrows in Fig. 1). This feature suggests that this young star, the most massive in the region, is the dominant source of the large CP. No clear pattern is observed around BN, although some negative CP could also be associated with this source. A quadrupolar CP distribution around a YSO is expected from the scattering observed in nebulae with bipolar cavities (Gledhill et al. 1996; Chrysostomou et al. 2000), as shown in numerical simulations (Whitney & Wolff 2002). Therefore, there may exist an embedded bipolar cavity along the northwest to southeast (Minchin et al. 1991). Interestingly, the central axis of the quadrupolar pattern or bipolar cavity appears to be aligned with the supposed direction of the magnetic field lines (Schleuning 1998; Kusakabe et al. 2008).

3.2. *Correlations among CP, LP, and H-$K_s$ Color*

Here, we investigate the origin of CP of the Orion star-forming region, using a detailed study of the correlation of our circular polarimetry and linear polarimetry results. First, a relatively good correlation between the degree of CP and the degree of LP in the $K_s$ band is shown in Figure 2*a*. The amplitude of the degree of the CP increases with an increase in the degree of LP. This finding agrees with the tendency shown by Chrysostomou et al. (2000). However, this can be consistent with all three candidates for CP mechanisms. Although several mechanisms can occur simultaneously, CP and extinction are well correlated in this region as Figure 1*b* shows. Therefore, we investigate the possibility of dichroic extinction being the main mechanism for CP in the Orion region.

When dichroic extinction by aligned axisymmetric grains is effective, a simple relationship can be derived. For production of CP under this assumption, Stokes $U$ is converted to Stokes $V$ by birefringence. According to previous studies (Martin 1974; Mishchenko 1991; Whitney & Wolff 2002), Stokes parameters of an incident photon (denoted by subscript inc) is modified into new Stokes parameters (denoted by subscript ext) by dichroic extinction:

$$U_{ext} = \exp(-n\,K_{11}\,s)\{U_{inc}\,\cos(n\,K_{34}\,s) - V_{inc}\,\sin(n\,K_{34}\,s)\}, \quad (1)$$

$$V_{ext} = \exp(-n\,K_{11}\,s)\{V_{inc}\,\cos(n\,K_{34}\,s) + U_{inc}\,\sin(n\,K_{34}\,s)\}. \quad (2)$$

The equations are valid for light rays traversing a distance $s$ through a uniform medium in which the number density of dust grains is $n$. The $K_{ij}$ terms are the independent term in the extinction matrix appropriate to the medium at the wavelength of observation. Focusing on the production of CP, we assume that $V_{inc}$ is zero. For example, first scattered photons by the outflow cavity after ejection of natural photons from the central star have no CP under Mie scattering, or only low CP if scattered off weakly aligned grains - as might occur for dust in the cavity wall. Thus, we obtain

$$V_{ext} = U_{ext}\,\tan(n\,K_{34}\,s). \quad\quad\quad (3)$$

When the tangential term $\tan(n\,K_{34}\,s)$ varies little, the correlation between Stokes $V$ and $U$ might indicate the value of $\tan(n\,K_{34}\,s)$; the term $n\,K_{34}\,s$ is the product of $(n\,K_{11}\,s)\,(K_{34}/K_{11})$, which are related to the optical depth and average dust property in the region.

To further simplify equation (3), we replace the tangential term, $\tan(n\,K_{34}\,s)$, with an approximation, because Stokes $V$ is much smaller than Stokes $U$ over almost all areas in our dataset (indicated by Fig. 3$b$; |V/U| < 0.1). Therefore, we can approximate equation (3)

as

$$V_{ext} \sim U_{ext} \left( n\,K_{34}\,s \right) = U_{ext}\,n\,K_{11}\,s\,\left( K_{34}\,/\,K_{11} \right). \qquad (4)$$

The term $n\,K_{11}\,s$ approximately corresponds to optical depth (Whitney & Wolff 2002). When we assume that the optical depth term $n\,K_{11}\,s$ is estimated by Av, and that the dust property $K_{34}\,/\,K_{11}$ is homogeneous in the observed region, the term $n\,K_{11}\,s\,(K_{34}\,/\,K_{11})$ is proportional to Av. It seems reasonable to assume that dust properties and grain alignment are homogeneous as a first approximation in the observed region (Schleuning 1998; Kusakabe et al. 2008), although slight changes in the dust properties in a single dark cloud are suggested from the extinction law (e.g., Naoi et al. 2007). If we estimate Av by the $H$-$K_s$ color, equation (4) can be rewritten as

$$V_{ext} \sim U_{ext} \left( H - K_s \right) \times \text{const.} \qquad (5)$$

These Stokes $V_{ext}$ and $U_{ext}$ in equations (3) and (5) are observed. It should be noted that equations (3) and (5) do not include Stokes $Q_{ext}$. Thus, a simple relationship for dichroic extinction is obtained.

The Stokes parameters described above are defined relative to the axis of grain alignment (Whitney & Wolff 2002), which is parallel to positive Stokes $Q$. To compare the above equations with the present observations, we must rotate the Stokes parameters in the observational frame into the grain alignment frame (only $Q$ and $U$ change with rotation). The grain alignment axis depends on the direction of the magnetic field. LP vectors of point sources by Kusakabe et al. (2008) are well-aligned in the direction of northwest to southeast, projected onto the sky. The vectors are thought to trace the magnetic field direction in the OMC-1 region. Therefore, we can assume that the direction of the magnetic field in this region is globally homogeneous. We take the position angle of the alignment axis as 120°, which is consistent with the inferred direction in previous observations (Schleuning 1998; Chrysostomou et al. 2000).

As is inferred from equation (5), extinction (e.g., estimated by $H$-$K_s$) can be correlated with CP caused by dichroic extinction. Figure 1$b$ indicates that CP is well-correlated with extinction. Furthermore, the correlation is more clearly shown by a pixel-by-pixel correlation between Stokes $V$ and $H$-$K_s$ color in Figure 2$b$. CP expressed by Stokes $V$ increases with extinction expressed by the $H$-$K_s$ color.

The pixel-by-pixel correlations between Stokes $V$ and $Q$, and Stokes $V$ and $U$ in the $K_s$ band are shown in Figures 3$a$ and 3$b$, respectively. Stokes parameters in reference to the axis of P.A. 120° are used. Stokes $V$ and $Q$ show no significant correlation in Figure 3$a$. In contrast, there is a strong linear relationship between Stokes $V$ and $U$ in Figure 3$b$. The correlation coefficients between Stokes $V$ and -$Q$, and $V$ and -$U$ are estimated to be about 0.2, and 0.6, respectively, when we include only data with $-5 \leq V \leq 10$ and $-40 \leq Q$ or $U \leq 40$ (ADU sec$^{-1}$ pixel$^{-2}$) to have less contamination by point sources, and exclude the data below 3σ to suppress the contributions from sky background noise (1σ~1.2 ADU sec$^{-1}$ pixel$^{-2}$). If most of the CP is produced by dichroic extinction, as opposed to scattering, then according to equation (3), Stokes $V$ should be correlated with Stokes $U$, especially when the tangential term tan($n$ $K_{34}$ $s$) varies little. Furthermore, Figure 3$c$ shows the pixel-by-pixel correlation between Stokes $V$ in the $K_s$ band and the product of Stokes $U$ and $H$-$K_s$ color. The correlation coefficient between $V$ and $-U(H$-$K_s)$ is estimated to be about 0.6, under the same condition (data within $-5 \leq V \leq 10$ and $-40 \leq U(H$-$K_s) \leq 40$ excluding the data below 3σ). Figure 3$c$ shows a relatively good linear relationship, as was suggested by equation (5).

These results suggest the effectiveness of dichroic extinction in this observed region.

There are several branches in the correlation diagrams in Figure 3. Figures 3*b* and 3*c* show one faint branch that extends to large *V* around small *U*, or *U* (*H-K$_s$*) which is caused by small *U* inferred from Figures 1*b* and 2*b*. This branch corresponds to point sources and low signal-to-noise ratio pixels. Deviations from the correlation may be caused by the assumption of the homogeneous direction of grain alignment. Indeed, the magnetic field line in the bright CP region (east from IRc2) may be inclined about 30° from the currently assumed direction (Chrysostomou et al. 1994). Numerical calculations for this region would be useful to study further (Chrysostomou et al. 2007).

In summary, our observations show good correlations between CP (i.e., Stokes *V*) and *H-K$_s$* color as a proxy for extinction, and between CP and LP oriented at P.A. ±45° relative to the grain alignment axis inferred from the magnetic field (i.e., Stokes *U*, or *U*(*H-K$_s$*) including the color). By contrast, the data show a lack of correlation between CP and LP parallel or perpendicular to the alignment axis (i.e., Stokes *Q*). These results indicate the major contribution of dichroic extinction to production of CP in this massive star-forming region. The quadrupolar pattern of the infrared reflection nebula around IRc2 can be basically explained in terms of a combination of scattering in a bipolar nebula and parent cloud extinction. In addition, our study indicates that aligned non-spherical grains exist widely in this massive star-forming region.

We thank the referee, Michael Wolff, for his very insightful comments and helpful suggestions. We acknowledge discussions with T. Nagayama and S. Sato. Thanks are due to the staff in SAAO. T.F. is supported by Research Fellowships of the Japan Society for the Promotion of Science for Young Scientists. M.T. is supported by Grants-in-Aid from the MEXT (16077101, 16077204), and that from the JSPS (19204018). This work was partially supported by KAKENHI 18-3219 and the Grant-in-Aid for the GCOE Program "The Next Generation of Physics, Spun from Universality and Emergence" from the MEXT of Japan.

*Facilities:* IRSF (SIRIUS, SIRPOL)

Fig. 1.—(*a*) Image of the degree of CP (%) in the $K_s$ band (2.14 μm) of the BN/KL region. The field-of-view is 2.7′ square. North is up and east is to the left. A positive sign for CP indicates that the electric vector is rotated anticlockwise in a fixed plane relative to the observer. IRc2 is indicated by arrows. BN is the red dot located at R.A. = $05^h35^m14^s.117$, decl. = -05°22′22.90″ (J2000). (*b*) *H-$K_s$* color image overlaid with contours of the corresponding degree of CP in the $K_s$ band. The contours start from −4.5% to −0.5% with an interval of 1% for negative CP (red lines), and from 0.5% to 16.5% with an interval of 1% for positive CP (white lines).

Fig. 2.—Pixel-by-pixel correlation of the BN/KL region using logarithmic color: (*a*) between the degree of CP and the degree of LP in the $K_s$ band; (*b*) between Stokes *V* in the $K_s$ band and *H-$K_s$* color. In each diagram, counts of sources which have a certain CP and LP, or Stokes *V* and *H-$K_s$* color are expressed. Before these correlation analyses, regions without sufficient signals (-0.5% < *P* < 1%) on the low-pass filtered CP image (after median smoothing over 13×13 pixels) are masked to have sufficient S/N ratios and less contamination by point sources.

Fig. 3.—Pixel-by-pixel correlation of the BN/KL region using logarithmic color: (*a*) between Stokes $V$ and $Q$ in the $K_s$ band; (*b*) between Stokes $V$ and $U$ in the $K_s$ band; (*c*) between Stokes $V$ in the $K_s$ band and the product of Stokes $U$ and $H$-$K_s$ color. In each diagram, counts of sources that have certain Stokes Parameters or Stokes $V$ and the product $U$ ($H$-$K_s$) are expressed. Stokes parameters have been converted based on the inferred axis of grain alignment (P.A. 120°). The sign of the horizontal axis is inverted.

**Fig. 1**

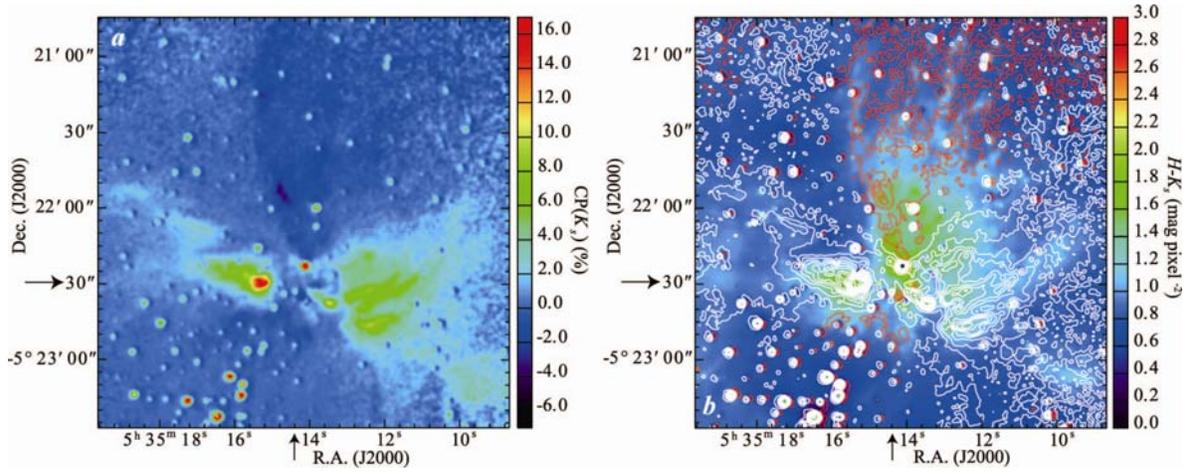

**Fig. 2**

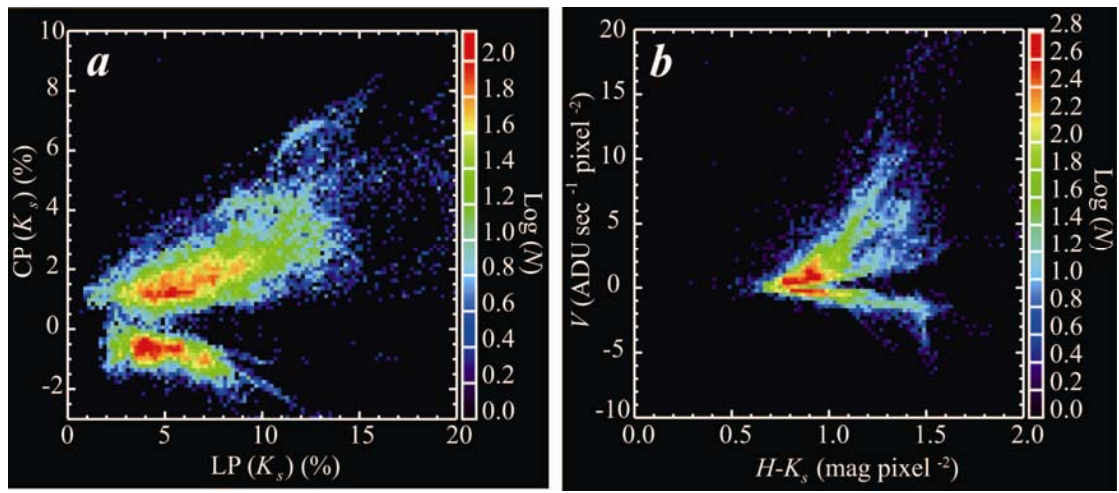

**Fig. 3**

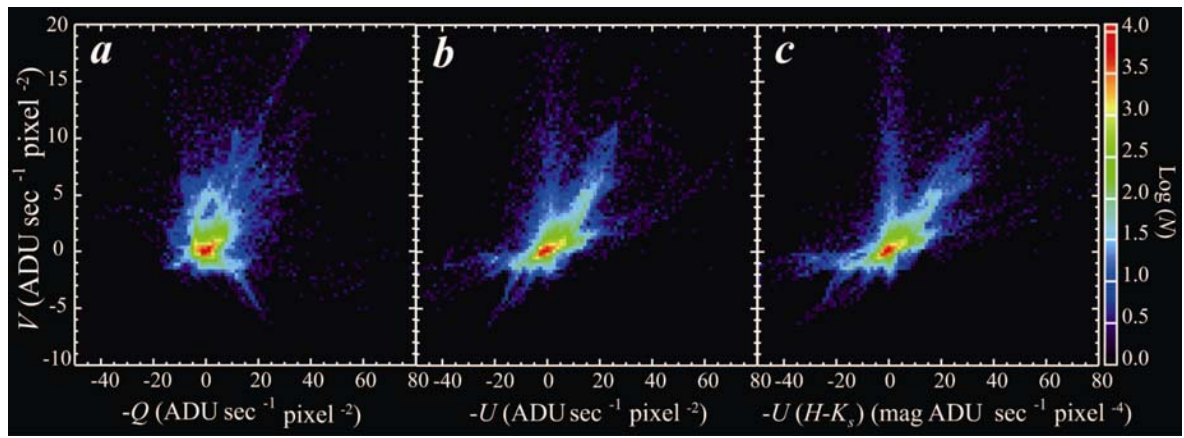